\documentclass[12pt,a4paper]{iopart}

\usepackage{iopams}
\usepackage{graphicx}

\usepackage{color}

\begin{document}

\title[Effective Langevin equations for a polar tracer in an active bath]{Effective Langevin equations for a polar tracer in an active bath}

\author{Milo\v s Kne\v zevi\'c and Holger Stark}

\address{Institut f\"ur Theoretische Physik, Technische Universit\"at Berlin, Hardenbergstra\ss e 
36, D-10623 Berlin, Germany}
\ead{knezevic@campus.tu-berlin.de}
\vspace{10pt}
\begin{indented}
\item[]August 2020
\end{indented}

\begin{abstract}
We study a polar tracer, having a concave surface, immersed in a two-dimensional suspension of active particles. Using Brownian dynamics simulations, we measure the distributions and auto-correlation functions of forces and torque exerted by active particles on the tracer. The tracer experiences a finite average force along its polar axis, while all the correlation functions show exponential decay in time. Using these insights we construct the full coarse-grained Langevin description for tracer position and orientation, where the active particles are subsumed into
an effective self-propulsion force and exponentially correlated noise. The ensuing mesoscopic dynamics can be described in terms of five dimensionless parameters. We perform a thorough parameter study of the mean squared displacement, which illustrates how the different parameters influence the tracer dynamics, which crosses over from a ballistic to diffusive motion. We also demonstrate that the distribution of tracer displacements evolves from a non-Gaussian shape at early stages to a Gaussian behavior for sufficiently long times.
\end{abstract}

%
%
\submitto{\NJP}
%
\maketitle
%
%

\section{Introduction}
In recent years active motion has evolved into a thriving field combining different disciplines from physics and chemistry to biology and engineering sciences~\cite{cates15,elgeti15,bechinger16,zoettl16}. Microorganisms swim in a fluid environment at low Reynolds number, meaning that viscous forces dictate over inertial forces. Tremendous research activities have been devoted to better understand their propulsion mechanisms~\cite{elgeti15,lauga09,ali16}, as well as to construct artificial microswimmers~\cite{deseigne10,theurkauff12,buttinoni13,palacci13} and to explore their fascinating patterns of collective motion~\cite{schaller10,wensink12,bricard13,zoettl14,pohl14}. Artificial or biological microswimmers, which we simply term active particles, consume energy to swim forward, and therefore are constantly driven out of equilibrium. Fascinating generic properties arise in such nonequilibrium settings, as illustrated, for example, by active particles getting stuck at confining walls~\cite{elgeti09,li09,elgeti13,schaar15}, on which they exert a swim pressure~\cite{takatori14,solon15n,solon15,fily18,zakine20}.

Combining active motion with concepts from Brownian ratchets, one of the existing paradigms in nonequilibrium statistical mechanics~\cite{reimann02}, provides new possibilities of rectified motion~\cite{reichhardt17}. In the direction of applications the following works are of interest:
capturing active particles~\cite{kaiser12,kaiser13}, sorting active particles based on their velocity~\cite{mijalkov13} or the mechanism how they reorient \cite{khatami16}, effective interactions between inclusions in active suspensions~\cite{angelani11prl,ray14,harder14,ni15,baek18,knezevic19}, cargo transport~\cite{palacci13jacs,koumakis13} and active assembly~\cite{simmchen15,stenhammar16,mallory18}. 
Active particles accumulate in corners, which causes directed transport through a wall of funnels~\cite{galajda07,wan08}, in an asymmetric potential~\cite{angelani11,fiasconaro08}, or in a symmetric potential in combination with a position-dependent swimming speed~\cite{pototsky13}, in a corrugated channel~\cite{ghosh13}, and in arrays of asymmetric obstacles~\cite{reichhardt13}.

When many active particles act on a mesoscopic object, they can be regarded as a nonequilibrium {\it active bath}, which is strongly determined by fluctuations in the swimming directions of particles. Rotational and translational ratchet motors can be constructed by placing asymmetric objects in active baths. Notably, a wheel with sawtooth-like contour deposited in an active bath exhibits unidirectional rotation~\cite{angelani09,dileonardo10,sokolov10}. When passive mesoscopic
objects, which do not self-propel, are suspended in such a bath, they are stochastically pushed around, and, more importantly, their motion can even be rectified if they have a polar shape and a pronounced concave surface~\cite{angelani10}. In what follows we shall refer to such a mesoscopic object as a polar tracer. Well known examples include semicircle forms and wedge-like structures~\cite{angelani10,kaiser14,mallory14}. The directed motion of the tracer can be explained by the fact that a portion of active particles, trapped within some cavity of the object, exercise certain pressure on the surface of the cavity and thus they push the object in the outward direction as illustrated in figure~\ref{fig1}. Thereby, the polar tracers are endowed with substantial persistence of motion and can act as microshuttles~\cite{angelani10,kaiser14,mallory14}.
Contrarily, spherical tracers in an active bath display only enhanced diffusive motion~\cite{wu00,gregoire01,chen07,leptos09,mino11,valeriani11,morozov14,maggi14}.

Most theoretical studies~\cite{angelani10,kaiser14,mallory14} performed so far were based on methods of Brownian dynamics simulations, replicating a collection of active particles interacting with the tracer. An alternative approach is the extraction of effective Langevin equations from 
the microscopic many-particle dynamics, which is a long term goal of theoreticians both in and out of thermal equilibrium.
Specifically, for an active bath it remains a challenge to describe the motion of a polar tracer or of even more complicated structures by mesoscopic equations. Due to the nonequilibrium nature of the bath~\cite{maes14}, it is clear that the standard Langevin equation is not appropriate in this case. It has been shown, for instance, that the motion of a spherical tracer in a bath of {\it E. coli} bacteria can be described by a Langevin equation containing instantaneous friction kernel and colored noise~\cite{wu00,angelani10,maggi14,maggi17}. Such noise can be generated by an auxiliary Ornstein--Uhlenbeck process~\cite{uhlenbeck30} and it brings the system outside of thermodynamic equilibrium.

In this article we develop an effective Langevin description for a polar tracer with a concave surface immersed in an active bath (see figure~\ref{fig1}). To determine the coarse-grained active noise resulting from the impact of the active bath particles, we performed simulations based on Brownian dynamics equations by extending previous studies. Our simulations support previous findings~\cite{angelani10,mallory14} concerning the existence of a finite average force acting along tracer's symmetry axis and the exponential time decay of relevant correlation functions. In addition, we demonstrate that the cross-correlation function between the torque and the force 
acting perpendicularly on the tracer's symmetry axis is always negative, but decays exponentially with time as well. Based on these insights we propose a complete description of the tracer motion with effective Langevin equations. More precisely, we show that the previous complex problem of
many-body Brownian dynamics can be reduced to a simple system of three stochastic equations of Langevin type. Using this approach, we performed a detailed study of the tracer mean squared displacement and its displacement probability distributions as a function of time. We show for the first time that the distribution of tracer displacements crosses over from a non-Gaussian at early stages of evolution to a Gaussian behavior for sufficiently long times.

The article is organized as follows: After the introduction, section~\ref{model} is devoted to the presentation of our model and its description in the frameworks of Brownian dynamics and effective Langevin equations approach. In section~\ref{results} we present our results together with an extensive discussion of the mean squared tracer displacement and associated probability distributions. Some concluding remarks and a summary of the main results are given in section~\ref{conclusion}. Finally, in~\ref{appA} we describe some technical details of 
our Brownian dynamics simulations.   

\section{Model}\label{model}

\begin{figure}[htb]
	\begin{center}
		\includegraphics[width=14cm]{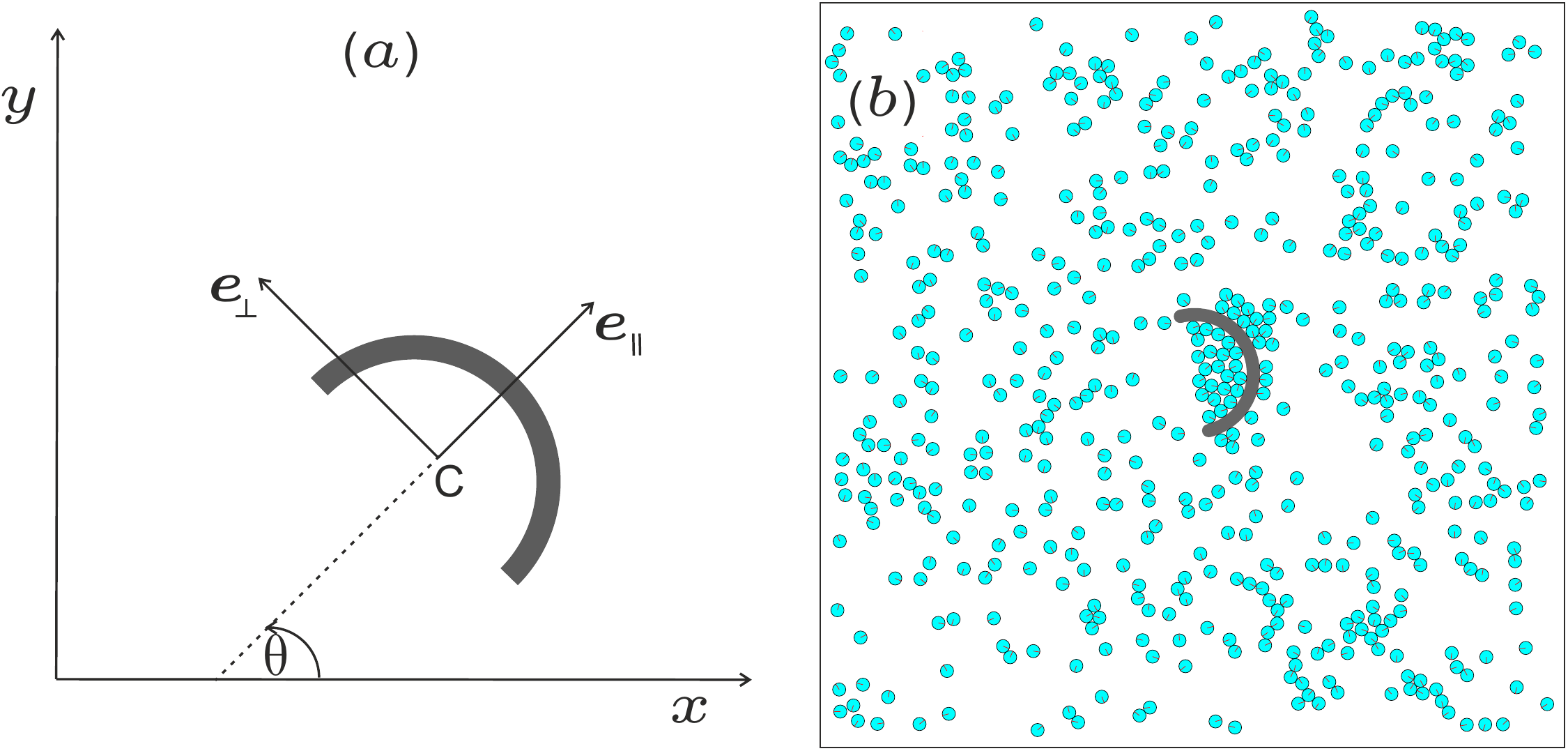}
		\caption{(a) Tracer geometry used in Brownian dynamics simulations. In the eigenframe of the tracer the orientation of its symmetry axis, passing through the center of mass C, is determined by the unit vector $\mathbf{e}_{\parallel}$ which makes the angle $\theta$ with the $x$-axis of the lab frame. (b) A snapshot of the motion of the tracer immersed in a bath of active particles; we selected only a small region of the simulation box centered around the tracer.}
		\label{fig1}
	\end{center}
\end{figure}

To introduce the quantities of interest for the coarse-grained dynamics, we start with a description of the problem within the framework of Brownian dynamics. Thus we begin with the general equations of tracer motion in the overdamped limit. In the lab frame, the vector $\mathbf{r} = (x,y)$ denotes the center of mass position of the tracer and the direction of its symmetry axis $\mathbf{e}_{\parallel}$ is characterized by an angle $\theta$ (see figure~\ref{fig1}). In dyadic notation the translational mobility matrix $\mathbf{M}$ of the tracer in its eigenframe can be written as
\begin{equation}
\mathbf{M} = \mu_{\parallel} \mathbf{e}_{\parallel} \otimes \mathbf{e}_{\parallel} + 
\mu_{\perp}(\mathbf{I}-\mathbf{e}_{\parallel} \otimes \mathbf{e}_{\parallel}),
\end{equation}
where $\mu_{\parallel}$ and $\mu_{\perp}$ are translational scalar mobilities, and $\mathbf{I}$ is the unit matrix. In the overdamped limit, where inertial contributions are negligible, the 
equations of motion of the tracer are
\begin{eqnarray}
\mathbf{V} &= \mathbf{M} \, \mathbf{F}, \label{velvec} \\
\dot\theta &= \kappa T \, .
\end{eqnarray}
Here, the tracer velocity $\mathbf{V}$ and the force $\mathbf{F}$ acting on it, in the eigenframe take the form $\mathbf{V} = v_{\parallel}\mathbf{e}_{\parallel} + v_{\perp}\mathbf{e}_{\perp}$ and $\mathbf{F} = F_{\parallel}\mathbf{e}_{\parallel} + F_{\perp}\mathbf{e}_{\perp}$, respectively.
The quantity $\dot\theta$ and $\mathbf{T} = T\mathbf{e}_z$ are the angular velocity of the tracer and the torque exerted by active particles on it, while $\kappa$ denotes its rotational mobility. For future convenience, we also introduce a typical length $l$ of the tracer connecting the mobilities $\mu_{\perp}$ and $\kappa$ through the relation $l^2=\mu_{\perp}/\kappa$.  

\begin{figure}[htb]
	\begin{center}
		\includegraphics[width=13.9cm]{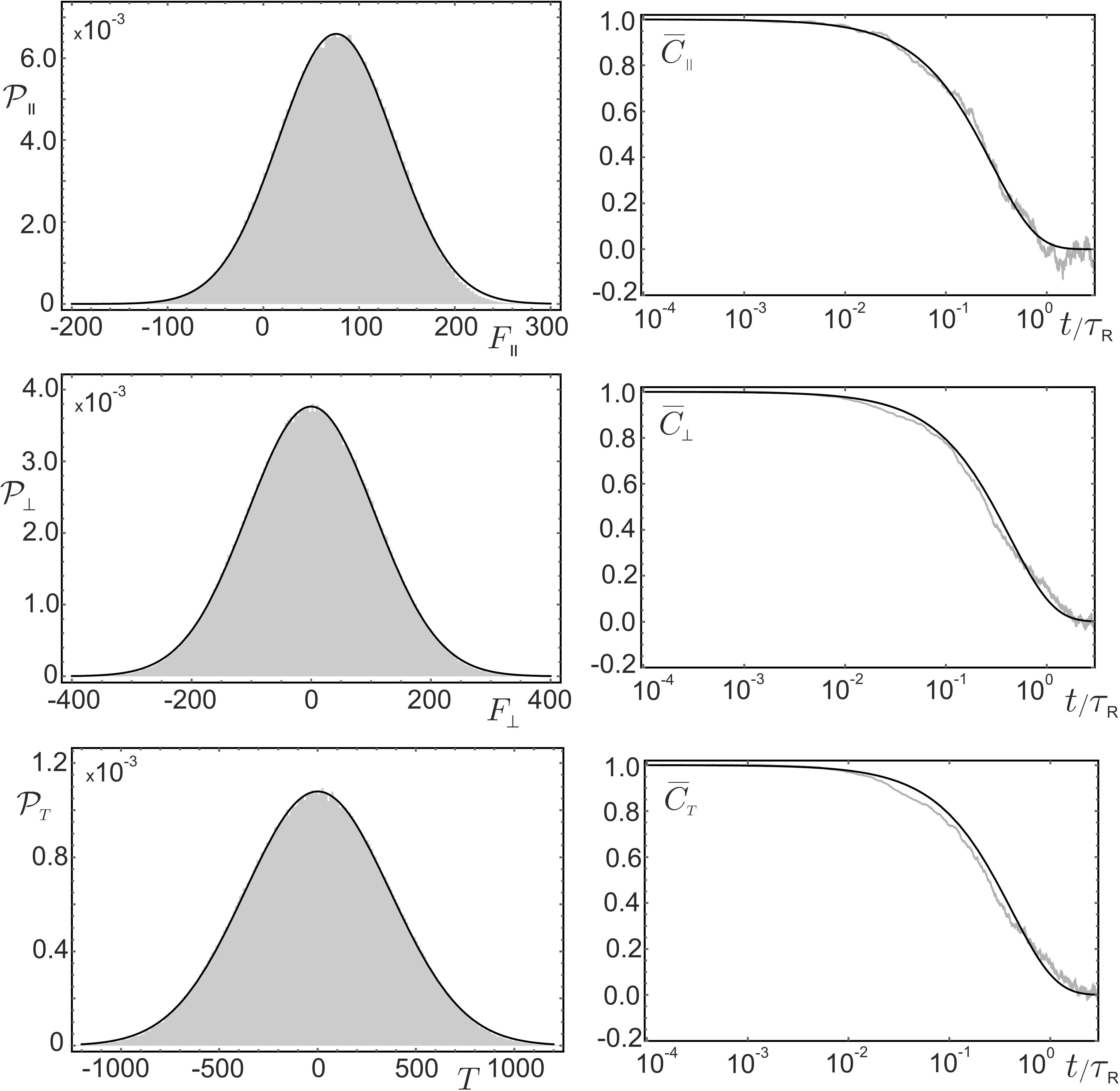}
		\caption{The probability distribution $\mathcal{P}_{\parallel}$ of the force $F_{\parallel}$ acting on the tracer is presented on the left side of the top row; the force
		is measured in units of $k_{\mathrm{B}}\mathrm{T}/\sigma$, where $\mathrm{T}$ is the temperature of the bath and $\sigma$ is the characteristic length of the interaction potential between the active particles. The solid black line is the best fit of $\mathcal{P}_{\parallel}$ to a Gaussian distribution; note that $\langle F_{\parallel} \rangle > 0$, which means that it has a positive projection on the polar axis
		$\mathbf{e}_{\parallel}$ of figure~\ref{fig1}. On the right side of the top row the scaled 
		auto-correlation function $\overline{C}_{\parallel}(t) = C_{\parallel}(t)/C_{\parallel}(0)$
		for $F_{\parallel}(t)$ is shown (light gray symbols), together with its fit
		to the exponential form $\rme^{-t/\tau_{\parallel}}$ (black solid line); note that the time
		is measured in units of the persistence time, $\tau_{\mathrm{R}}$, of an active particle, discussed in the~\ref{appA}. The middle row presents the corresponding data for $\mathcal{P}_{\perp} (F_\perp)$ and	$\overline{C}_{\perp}(t)$. Finally in the bottom row we presented the behavior of $\mathcal{P}_T$ and $\overline{C}_T$.
		Note that the mean values of the perpendicular force and torque acting on the tracer are vanishing, $\langle F_{\perp} \rangle = 0$ and $\langle T \rangle = 0$; the torque is measured in units of thermal energy $k_{\mathrm{B}}\mathrm{T}$. According to our observations the values of all three correlation times $\tau_{\parallel}$, $\tau_{\perp}$, and $\tau_T$ are approximately the same. All results presented in this panel correspond to an active bath having the area packing fraction $\phi \approx 0.08$ and the persistence number $P_{\mathrm{er}}=80/3$ of active particles (see~\ref{appA} for definitions of $\phi$ and $P_{\mathrm{er}}$ and more details).}
		\label{fig2}
	\end{center}
\end{figure}

To characterize the fluctuating force and torque, which result from the active particles hitting the polar tracer, we performed Brownian dynamics simulations of a semicircle tracer immersed in a bath of active Brownian particles~\cite{romanczuk12}. The technical details of simulations are given in~\ref{appA}. As one can infer from figure~\ref{fig2} (top row, left) the probability distribution of force $F_{\parallel}$ has a Gaussian profile centered at a finite mean (similar results were reported in~\cite{angelani10}, where the motion of a wedge shaped tracer in a bath of active rods has been studied). One can also see that the auto-correlation function $C_{\parallel}(t) = \langle F_{\parallel}(t_0)F_{\parallel}(t_0+t) \rangle - \langle F_{\parallel} \rangle^2$ in the stationary regime (top row, right) decays with time following an exponential law; similar behavior has been observed in~\cite{angelani10}. As the graph shows, the characteristic time of this decay is of the order of the reorientation time of an individual active bath particle. This hints that the force $F_{\parallel}$ can be modeled as $F_{\parallel} = \langle F_{\parallel} \rangle + \xi_{\parallel}$, where $\langle F_{\parallel} \rangle$ is a net drift force acting along the tracer's polar axis, and $\xi_{\parallel}$ is a random noise term, with a zero mean, exponentially correlated in time. In contrast to the case of $F_{\parallel}$, Brownian dynamics simulations show that the average value of the perpendicular component of the force is equal to zero, $\langle F_{\perp}\rangle = 0$, which suggests that the perpendicular force can be taken in the simple form $F_{\perp} = \xi_{\perp}$, with $\langle \xi_{\perp}\rangle=0$. As before, the corresponding correlation function $C_{\perp}(t)$ appears to follow an exponential decay in time (see figure~\ref{fig2}, middle row). Furthermore, the simulations also reveal that the different components of the random force are not mutually correlated, $\langle \xi_{\parallel}(t) \xi_{\perp}(t') \rangle = 0$, which is also clear by the polar symmetry of the tracer.
All these findings can be summarized by relations
\begin{equation}
\langle \xi_{\alpha} \rangle = 0, \quad \langle \xi_{\alpha}(t) \xi_{\beta}(t') \rangle = \delta_{\alpha \beta} \frac{1}{\mu_{\alpha}^2} D_{\alpha}^{\mathrm{A}} \frac{1}{\tau_{\alpha}}\rme^{-\frac{|t-t'|}{\tau_{\alpha}}},
\label{eqs.noise}
\end{equation}
where the indices $\alpha$ and $\beta$ take values $\parallel$ and $\perp$, while $D_{\alpha}^A$
are the diffusion constants along the principal directions and $\tau_{\alpha}$ represents the corresponding persistence time of the active noise.
\begin{figure}[htb]
	\begin{center}
		\includegraphics[width=11.5cm]{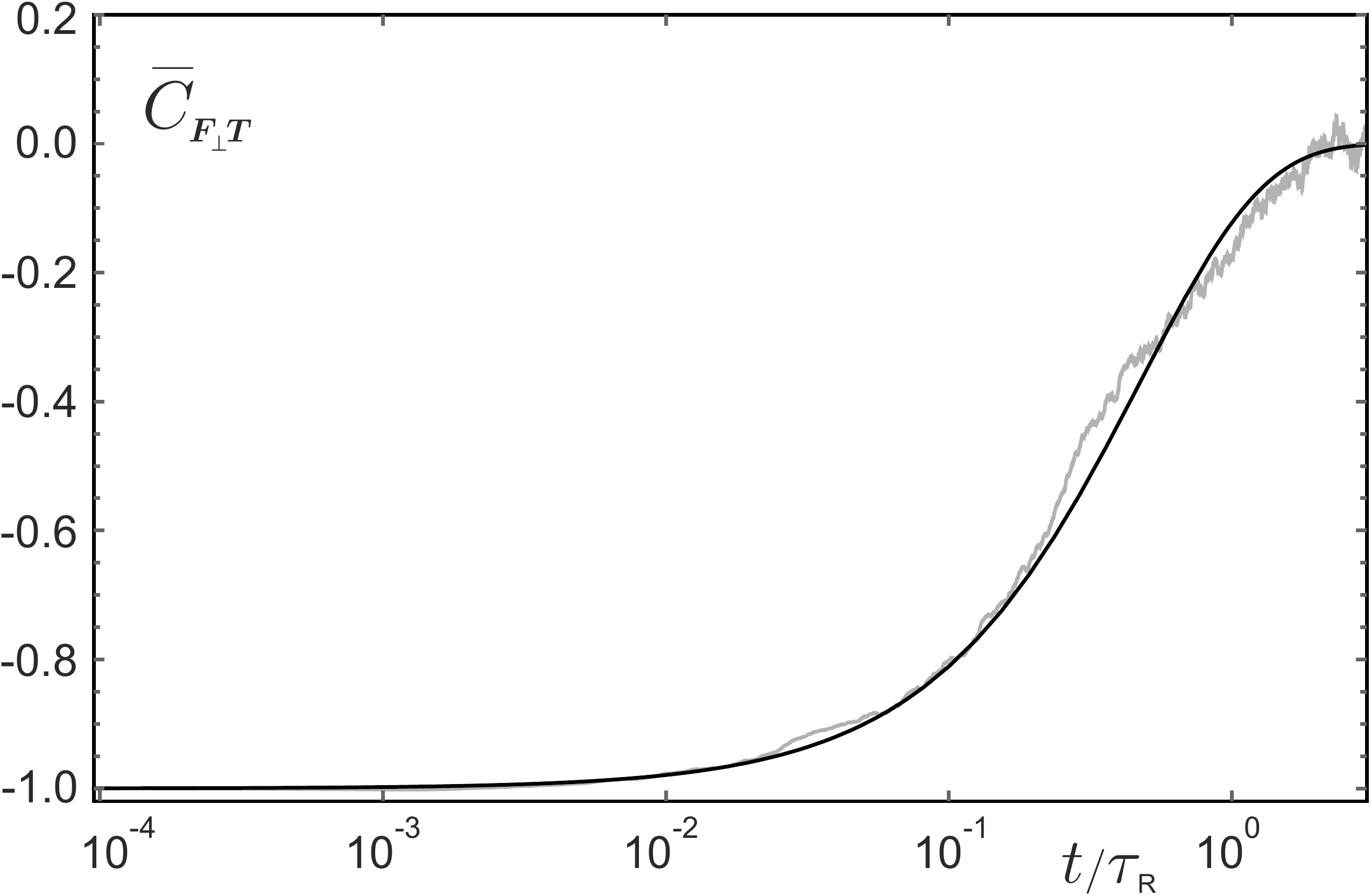}
		\caption{The scaled cross-correlation function between the perpendicular force and torque acting on the tracer, $\overline{C}_{F_{\perp}T}(t) = C_{F_{\perp}T}(t)/|C_{F_{\perp}T}(0)|$, versus time measured in units of $\tau_{\mathrm{R}}$ (light gray symbols); note that $F_{\perp}$ and $T$ are anti-correlated, $\overline{C}_{F_{\perp}T}<0$. The solid black curve represents the best fit of $\overline{C}_{F_{\perp}T}$ data to the exponential form $-\rme^{-t/\tau_{\mathrm{c}}}$, where $\tau_{\mathrm{c}}$ is the characteristic cross-correlation time. The simulation data were obtained for the same values of bath parameters as those in figure~\ref{fig2}.}
		\label{fig3}
	\end{center}
\end{figure}

We have seen that the active particles affect the motion of the tracer through the force $\mathbf{F}$. These particles also produce a random torque leading to a rotation of the tracer, which causes a gradual degradation in directed motion. Brownian dynamics simulations demonstrate that the average value of torque is equal to zero, and that its auto-correlation function $C_{\mathrm{T}}(t)$ decays exponentially in time (see figure~\ref{fig2} (bottom row) and~\cite{angelani10}). Since the fluctuating torque results from the fluctuating perpendicular force component, we also measured the cross-correlation function $C_{F_{\perp}T}(t) = \langle F_{\perp}(t_0) T(t_0+t) \rangle$. As one can see from figure~\ref{fig3}, it is always negative and displays an exponential behavior. In our coarse-grained model the presence of these anti-correlations is taken into account in the following way
\begin{equation}
T(t) = - \frac{\mu_{\perp}}{\kappa l_{\mathrm{T}}} \xi_{\perp}(t),
\end{equation}
where the characteristic length $\mu_{\perp}/(\kappa l_{\mathrm{T}})$ linking $T$ and $\xi_{\perp}$ can be deduced using dimensional analysis. In our numerical analysis it is however more convenient to work with the length $l_{\mathrm{T}}$ that we introduced in this proportionality factor. Let us mention in passing that, in contrast to the case of $F_{\perp}$, the force $F_{\parallel}$ and the torque $T$ are not mutually correlated.

Taking into account the above considerations, after transforming the equation (\ref{velvec}) into the lab frame, we obtain the following system of three stochastic equations for the polar tracer
\begin{eqnarray}
\dot x &= \mu_{\parallel}\left (\langle F_{\parallel} \rangle + \xi_{\parallel}\right )\cos\theta - \mu_{\perp}\xi_{\perp}\sin\theta, \label{xeq}\\
\dot y &= \mu_{\parallel}\left (\langle F_{\parallel} \rangle + \xi_{\parallel}\right )\sin\theta + \mu_{\perp}\xi_{\perp}\cos\theta, \label{yeq}\\
\dot \theta &= - \frac{\mu_{\perp}}{l_{\mathrm{T}}}\xi_{\perp}. \label{theta}
\end{eqnarray}
In these equations we neglected the usual thermal noise because we confined ourselves to the physically most interesting case of large speed of active particles. To generate the exponentially correlated noises $\xi_{\parallel}$ and $\xi_{\perp}$ of equations (\ref{eqs.noise}) with exactly the same parameters, we use two auxiliary Ornstein--Uhlenbeck processes~\cite{uhlenbeck30}
\begin{equation}
\dot \xi_{\alpha} = - \frac{1}{\tau_{\alpha}}\left (\xi_{\alpha} + \frac{\sqrt{2D_{\alpha}^A}}{\mu_{\alpha}} \eta_{\alpha}\right ) \quad \mathrm{with} \quad \alpha = \parallel, \perp,
\label{OU} 
\end{equation}
where $\eta_{\alpha}$ are Gaussian white noises of zero mean and unit variance: $\langle \eta_{\alpha}\rangle = 0$, $\langle \eta_{\alpha}(t)\eta_{\beta}(t') \rangle = \delta_{\alpha \beta} \delta(t-t')$. 

We use the typical extent of the tracer $l$ as the unit of length, persistence time $\tau_{\parallel}$ of the noise as the unit of time, and we measure forces in units of the effective self-propulsion force $\langle F_{\parallel} \rangle$. Now, keeping the same notation, the equations (\ref{xeq}) - (\ref{OU}) can be rewritten in the dimensionless form:
\begin{eqnarray}
\dot x = P \left [ (1+\xi_{\parallel})\cos\theta - \xi_{\perp}\sin\theta \right ], \label{xndeq}\\
\dot y = P \left [ (1+\xi_{\parallel})\sin\theta + \xi_{\perp}\sin\theta \right ], \label{yndeq}\\
\dot \theta = -\frac{P}{b}\xi_{\perp}, \label{thetand}\\
\dot \xi_{\parallel} = -\xi_{\parallel} + \frac{\sqrt{2Q}}{P}\eta_{\parallel}, \\ 
\dot \xi_{\perp} = -w \xi_{\perp} + \frac{w}{P} \sqrt{\frac{2Q}{d}} \eta_{\perp}, \label{xperpndeq} 
\end{eqnarray}
where we introduced five independent dimensionless parameters:
\begin{equation}
P = \frac{\mu_{\parallel}\langle F_{\parallel}\rangle \tau_{\parallel}}{l}, \quad
Q = \frac{D_{\parallel}^{\mathrm{A}} \tau_{\parallel}}{l^2}, \quad
d = \frac{D_{\parallel}^{\mathrm{A}}}{D_{\perp}^{\mathrm{A}}}, \quad
w = \frac{\tau_{\parallel}}{\tau_{\perp}}, \quad
b = \frac{l_{\mathrm{T}}}{l}. \label{params}
\end{equation}
Here, the persistence number $P$ quantifies the effective persistence length $\mu_{\parallel}\langle F_{\parallel}\rangle \tau_{\parallel}$, which is the distance the tracer traverses in roughly the same direction. The parameter $Q$ is the ratio of two timescales: the persistence time $\tau_{\parallel}$ and the time $l^2/D_{\parallel}^{\mathrm{A}}$ it takes the tracer to diffuse its own length $l$ due to active noise. The parameters $d$ and $w$ describe the ratios of diffusion constants and persistence times of the active noise along $\parallel$ and $\perp$ directions, respectively. Finally, $b$ denotes the characteristic length $l_{\mathrm{T}}$ that we introduced earlier measured in units of $l$. It is useful to note that all these dimensionless parameters depend on the geometry of the tracer and the active bath properties. Let us add yet that in writing the above dimensionless equations we removed the parameter $\mu_{\parallel}/\mu_{\perp}$ by absorbing it into the definition of the perpendicular component of noise: $\xi_{\perp}\mu_{\parallel}/\mu_{\perp} \rightarrow \xi_{\perp}$.

The stochastic equations (\ref{xndeq}) - (\ref{xperpndeq}) are integrated using a simple Euler scheme with a time step of $\delta t/\tau_{\parallel} = 10^{-5}$. The simulation time goes up to $t/\tau_{\parallel}=2000$, and the results are averaged over 100 independent simulation runs for each parameter set.

\section{Results}\label{results}

One of the most important characteristics of tracer movement is the behavior of its mean squared displacement (MSD), which will be presented in section~\ref{MSD}. Of course more detailed characterization of tracer's motion is provided by probability distributions of its displacements. We explore them in section~\ref{PDD}.

\subsection{Mean squared displacement}\label{MSD}

We analyze the motion of the tracer by computing its MSD:
$\langle \Delta\mathbf{r}^2(t) \rangle = \langle [\mathbf{r}(t+\mathcal{T}) - \mathbf{r}(\mathcal{T})]^2 \rangle_{\mathcal{T}}$, where the averaging is performed over different initial times $\mathcal{T}$ and over 100 independent simulation runs. Our results span over several decades in time. In the following we evaluate how the MSD changes with varying each of the above dimensionless parameters (\ref{params}).
\begin{figure}
	\begin{center}
		\includegraphics[width=13cm]{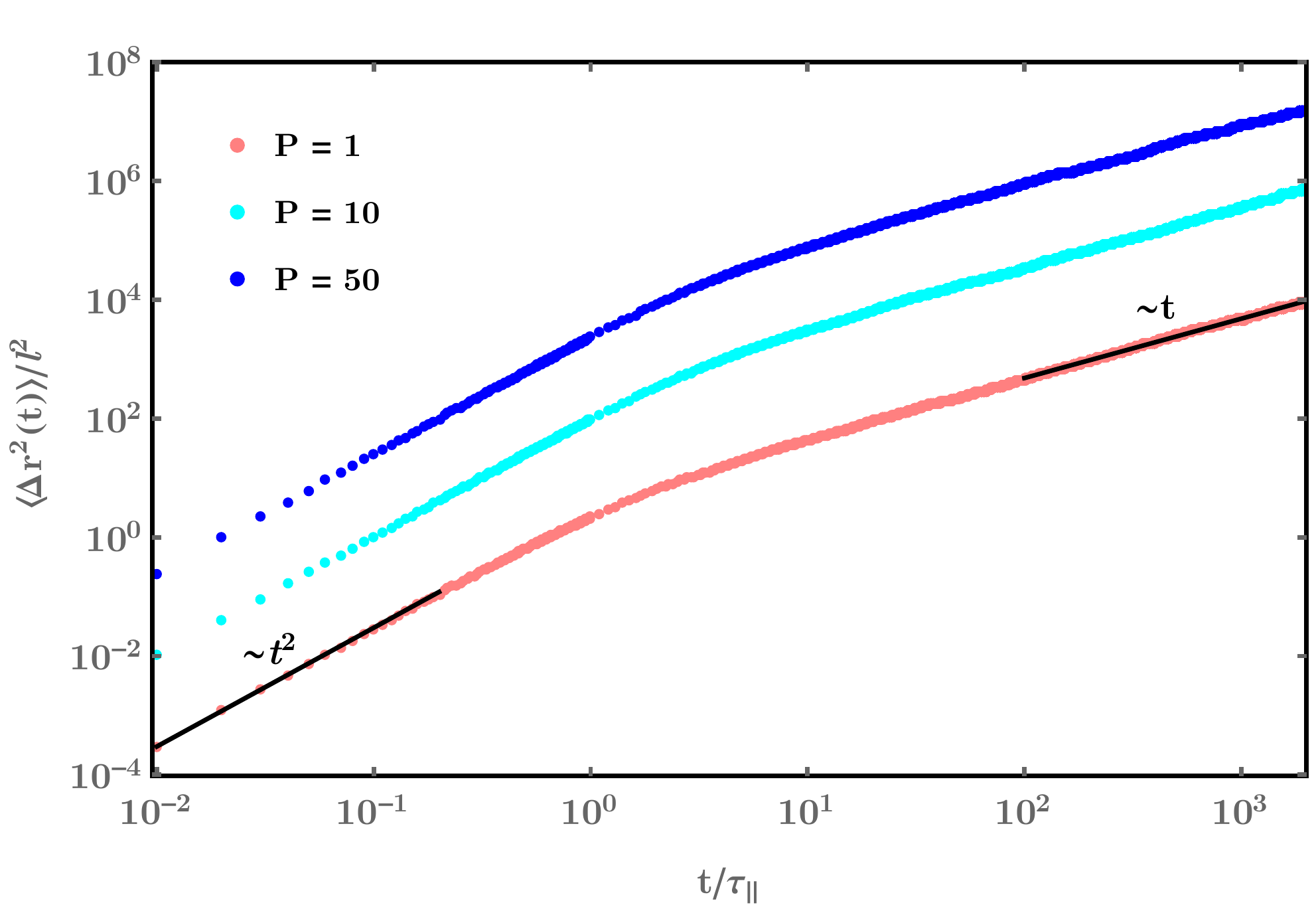}
		\caption{The MSD, measured in units of $l^2$, as a function of time, measured in units of $\tau_{\parallel}$, for three selected persistence number values $P$. All other dimensionless parameters are set to 1. The solid black lines are guides to the eye.		
		}
		\label{fig4}
	\end{center}
\end{figure}

The MSD obtained for different values of the persistence number $P$ is shown in figure~\ref{fig4}.
As one can infer from figure~\ref{fig4} the MSD displays a ballistic behavior, $\langle \Delta\mathbf{r}^2(t) \rangle \sim t^2$, for short times ($t/\tau_{\parallel}\lesssim1$, for our choice of parameters). The practically pure ballistic motion is due to the persistence in random force acting on the tracer (there are no thermal fluctuations in our model). The higher the persistence number, the more space is explored by the tracer. From the equations (\ref{xndeq}) and (\ref{yndeq}), it is easy to see that the MSD should scale as $\langle \Delta\mathbf{r}^2 \rangle \sim P^2 t^2$ in the ballistic regime, which is supported by the numerical results in figure~\ref{fig2}. On the other hand, for long times ($t/\tau_{\parallel}\gtrsim100$) the tracer motion is eventually randomized for all $P$ so that the normal diffusion sets in, $\langle \Delta\mathbf{r}^2(t) \rangle \sim t$. One can notice that a larger value of $P$ gives rise to an enhanced effective value of the diffusion coefficient.

\begin{figure}
	\begin{center}
		\includegraphics[width=13cm]{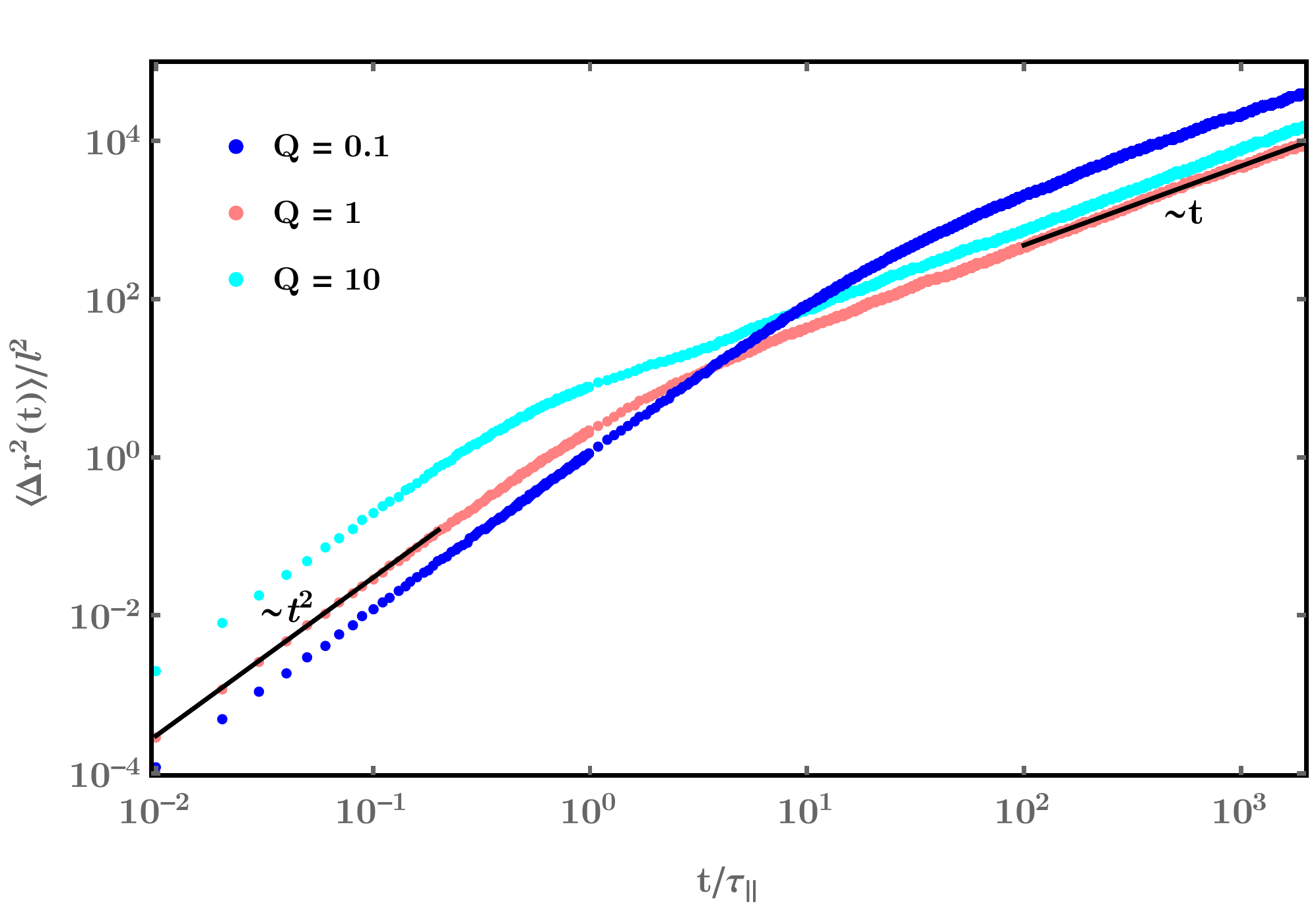}
		\caption{The MSD as a function of time for three selected values of parameter $Q$.
			All other dimensionless parameters are set to 1. The solid black lines are guides to the eye.}
		\label{fig5}
	\end{center}
\end{figure}

Varying parameter $Q = D_{\parallel}^{\mathrm{A}}\tau_{\parallel}/l^2$ yields a nontrivial change of the MSD, see figure~\ref{fig5}. By changing $Q$ one essentially alters the active diffusion constants $D_{\parallel}^{\mathrm{A}}$ and $D_{\perp}^{\mathrm{A}}=D_{\parallel}^{\mathrm{A}}/d$. Compared to the case $Q=1$, for $Q=10$ the tracer is subjected to a larger value of correlated noise, which also affects short-time ballistic motion leading to a larger effective speed. However, the tracer is also exposed to a greater active diffusion constant $D_{\perp}^{\mathrm{A}}$ or correlated noise along its lateral direction, causing a destruction of its ordered motion at earlier times if compared to the case $Q=1$. As a consequence of this, the effective diffusion constant at long times is not markedly distinct between these two cases. On the other hand, for $Q=0.1$ the lateral random force exerted on the tracer is sufficiently small, such that for a chosen $P=1$, one obtains a pronounced ballistic regime spanning up to times $t/\tau_{\parallel} \approx 10$. Consequently, the effective diffusion constant at long times is noticeably larger with respect to the previous two cases.
\begin{figure}
	\begin{center}
		\includegraphics[width=13cm]{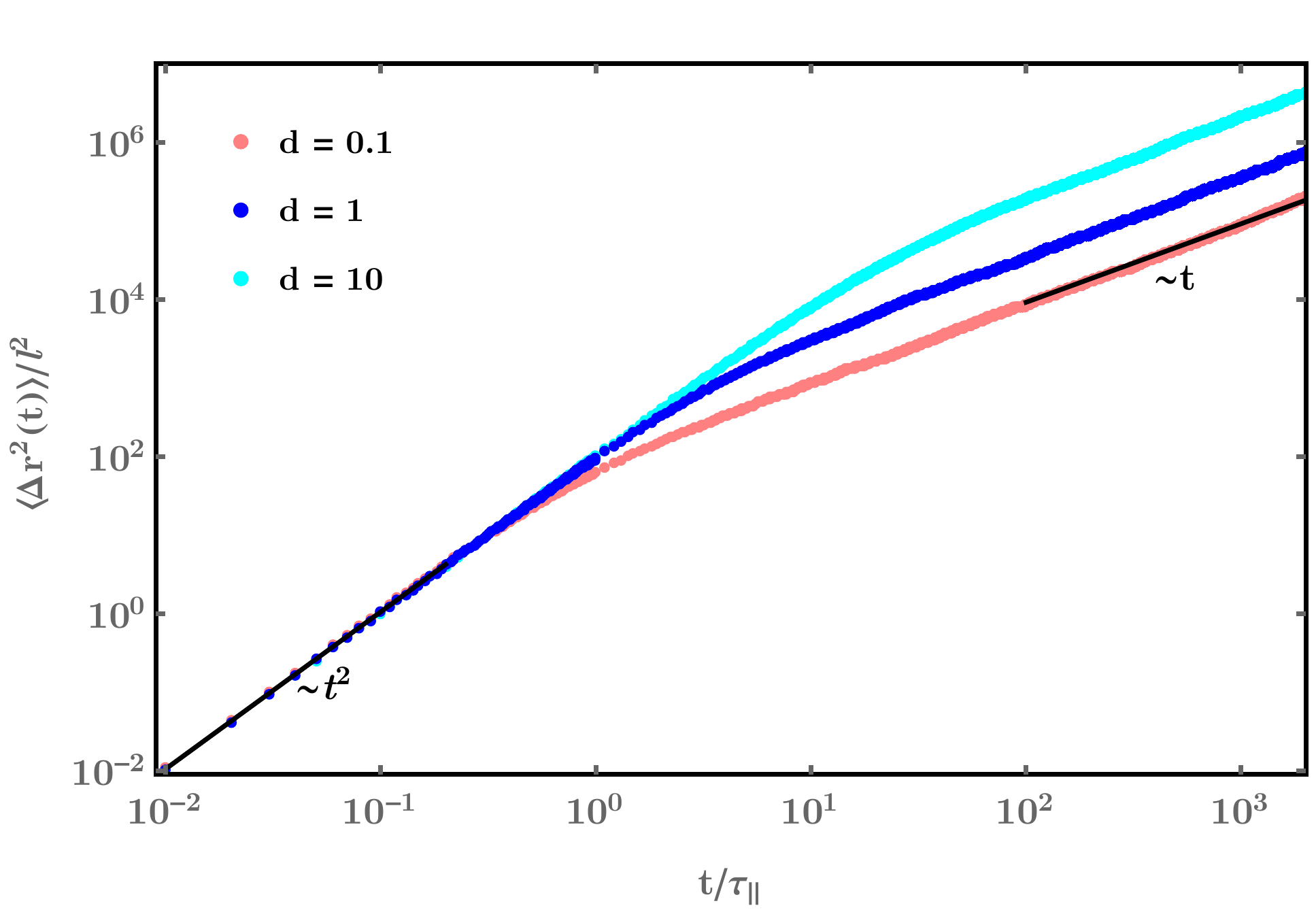}
		\caption{The MSD as a function of time for $P=10$ and three selected values of parameter $d$. All other dimensionless parameters are set to 1. The solid black lines are guides to the eye.}
		\label{fig6}
	\end{center}
\end{figure}
\begin{figure}[htb]
	\begin{center}
		\includegraphics[width=13cm]{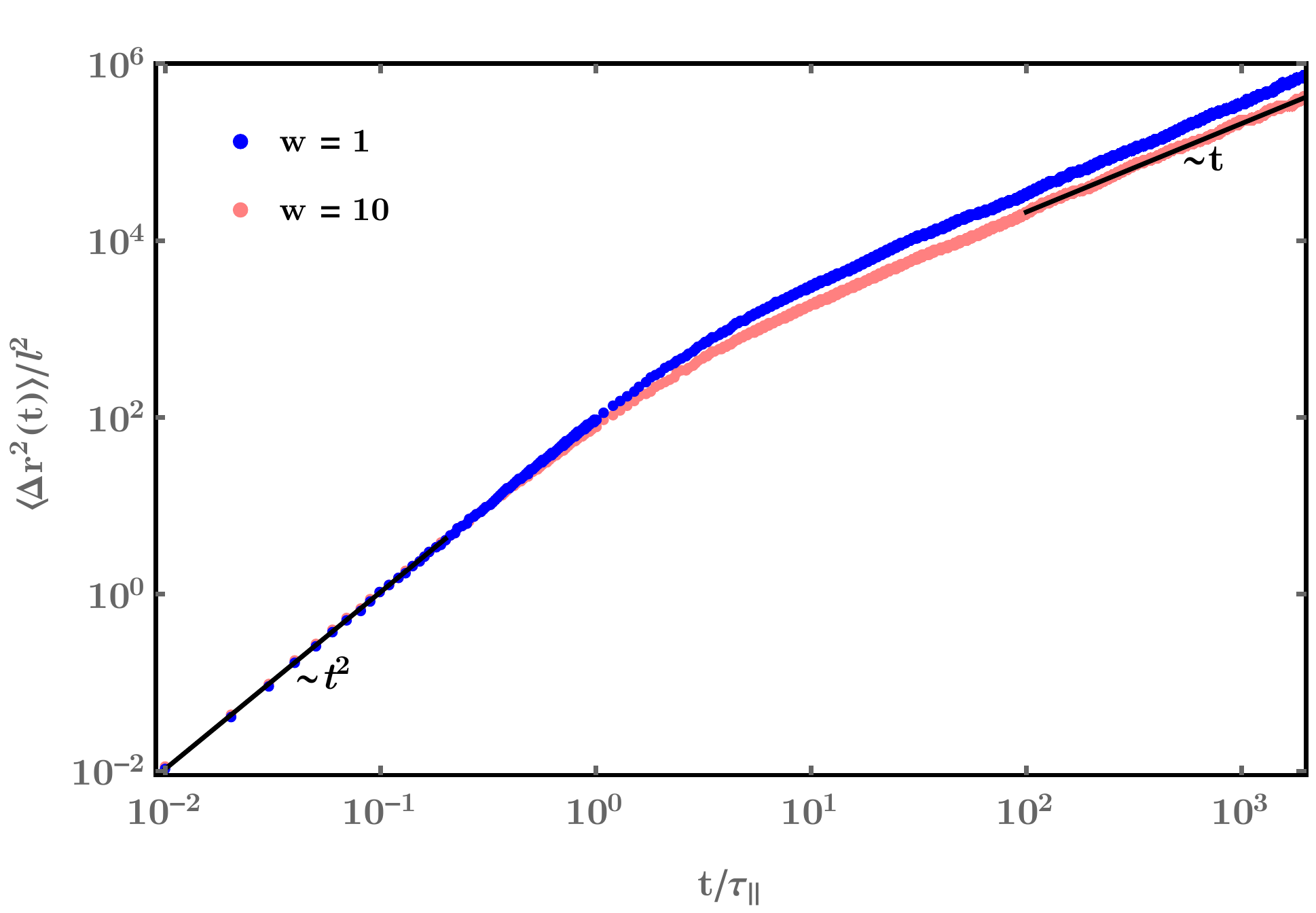}
		\caption{The MSD as a function of time for $P=10$ and $w=1,\ 10$. All other dimensionless parameters are set to 1. The solid black lines are guides to the eye.}
		\label{fig7}
	\end{center}
\end{figure}
\begin{figure}
	\begin{center}
		\includegraphics[width=13cm]{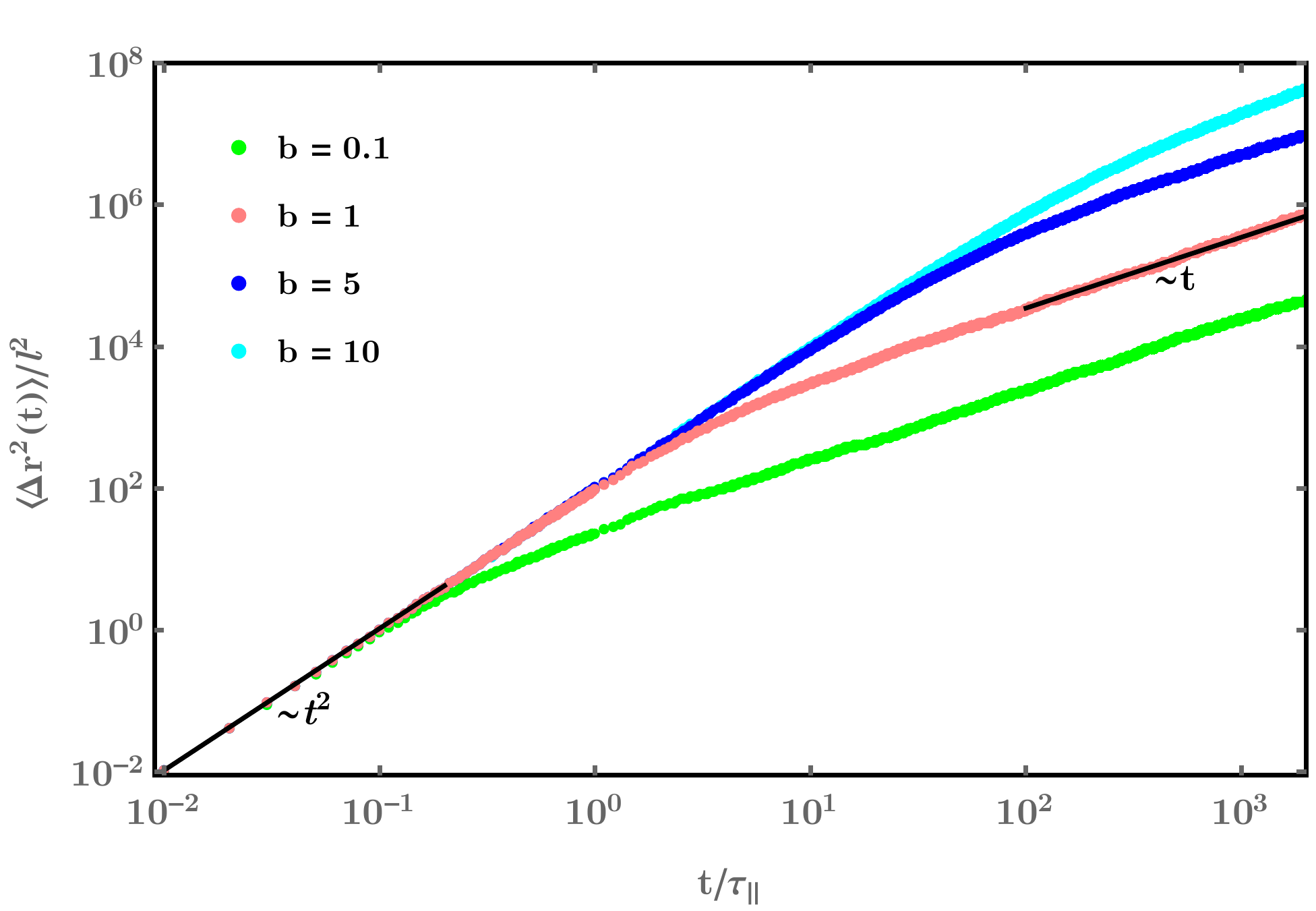}
		\caption{The MSD as a function of time for $P=10$ and four selected values of parameter $b$. All other dimensionless parameters are set to 1. The solid black lines are guides to the eye.}
		\label{fig8}
	\end{center}
\end{figure}

The MSD for persistence number $P=10$ and for diverse values of parameter $d$, quantifying the ratio of active diffusion constants along the main and lateral axis of the tracer, is shown in figure~\ref{fig6}. The effect of changing $d$ is straightforward. Increasing $d$ above the reference value $d=1$, corresponding to $D_{\parallel}^{\mathrm{A}}=D_{\perp}^{\mathrm{A}}$, the tracer exhibits longer ballistic movement due to elevated diffusion constant of the persistent active noise along its symmetry axis. In contrast, $d<1$ signifies less persistent ballistic motion. 

In figure~\ref{fig7} we show the MSD for $P=10$ and two values of the parameter $w=\tau_{\parallel}/\tau_{\perp}$. Note that the measurements of time auto-correlations of forces $F_{\parallel}$ and $F_{\perp}$ in Brownian dynamics simulations suggest that $\tau_{\parallel}>\tau_{\perp}$ with $w \gtrsim 1$. Thus, figure~\ref{fig7} indicates that in the physically relevant region of parameter space $1 \leq w < 10$ the MSD is not appreciably sensitive
to variations of $w$. This implies that in most practical cases one can set $w=1$. 

Finally, the effect of changing the parameter $b=l_{\mathrm{T}}/l$ on the MSD is depicted in figure~\ref{fig8}. As can be seen from the equation (\ref{thetand}) larger values of $b$ correspond
to a slower variation of tracer's angular velocity, and thus to a longer persistence of motion.
As before for early times we obtain a ballistic regime, while for longer times diffusive motion
takes place. One notes that the duration of the ballistic regime grows with $b$.

\subsection{Probability distribution of displacements}\label{PDD}
\begin{figure}[htb]
	\begin{center}
		\includegraphics[width=13cm]{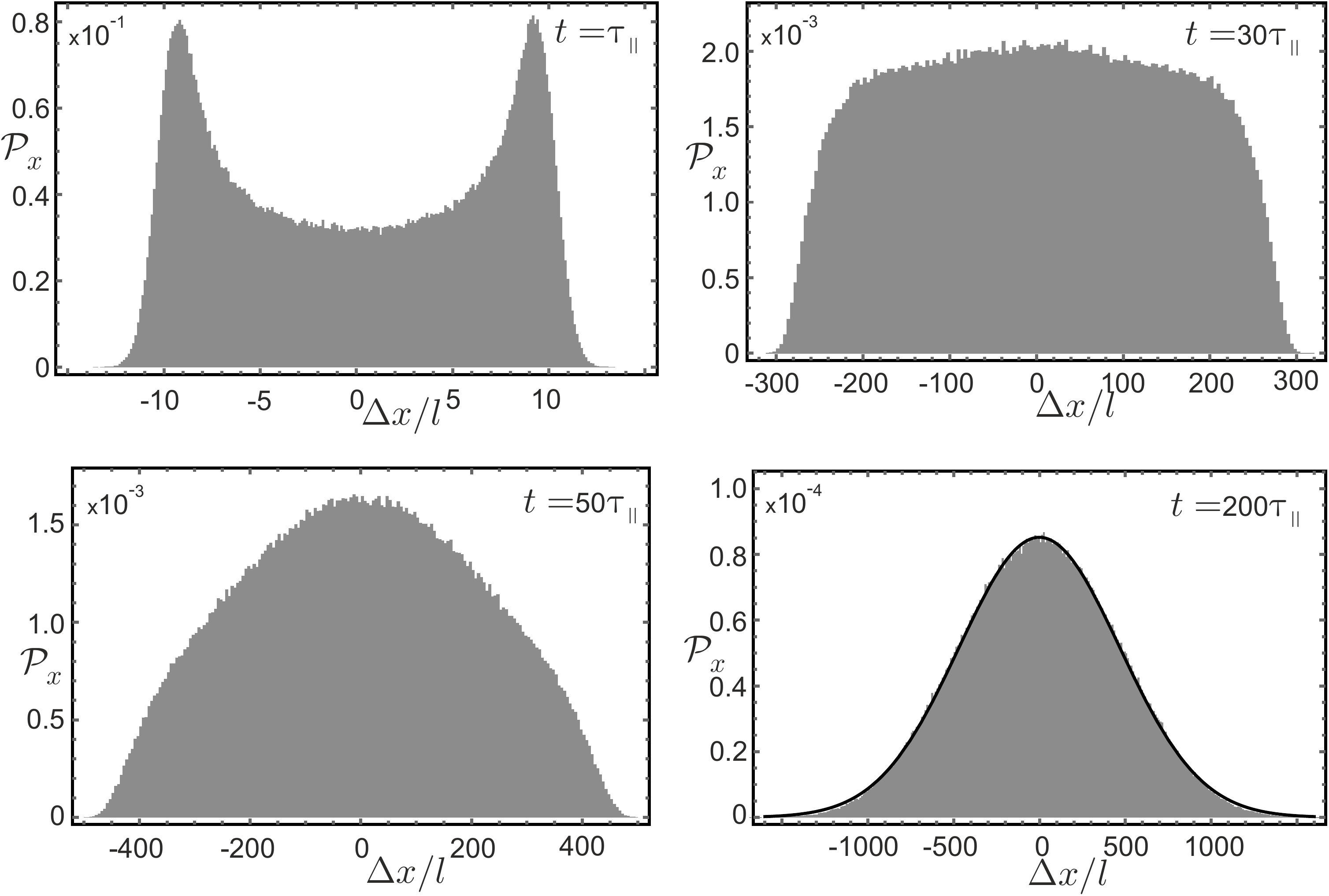}
		\caption{The probability distribution $\mathcal{P}_x$ of displacements $\Delta x$, measured
			in units of $l$, for four characteristic time values $t/\tau_{\parallel}$. Here we chose
			the same values of parameters as those used to obtain the cyan line in figure~\ref{fig6},
			$P=d=10$ and $Q=w=b=1$. For $t/\tau_{\parallel}=200$ the solid black line provides the best fit of $\mathcal{P}_x$ to a Gaussian form.}
		\label{fig9}
	\end{center}
\end{figure}

The time evolution of probability distribution $\mathcal{P}_x$ of tracer displacement $\Delta x = x-x_0$, with respect to some initial position $x_0$, obtained for some representative values of relevant parameters, is shown in figure~\ref{fig9}. 
As one can infer from this figure, at early times, when the tracer displays ballistic motion, the probability distribution $\mathcal{P}_x$ is bimodal with two peaks located at $\Delta x/l \approx \pm P$. As the time progresses, the height of these peaks decreases until the end of the ballistic
regime. After a characteristic time $t/\tau_{\parallel}$ (in our case $t/\tau_{\parallel}=30$) they
completely disappear, and $\mathcal{P}_x$ exhibits a plateau. Later in time (see the figure corresponding to $t/\tau_{\parallel} = 50$) the shoulders of $\mathcal{P}_x$ subside, and with further increase in time $\mathcal{P}_x$ crosses over to a purely Gaussian form for sufficiently long times. The width of this Gaussian is directly related to the MSD presented in figure~\ref{fig6}. The probability distributions $\mathcal{P}_x$ obtained from the stochastic equations (\ref{xndeq}) - (\ref{xperpndeq}) are in a good qualitative agreement with those retrieved from our Brownian dynamics simulations (see figure~\ref{figA1} of the~\ref{appA}).

\begin{figure}[htb]
	\begin{center}
		\includegraphics[width=13cm]{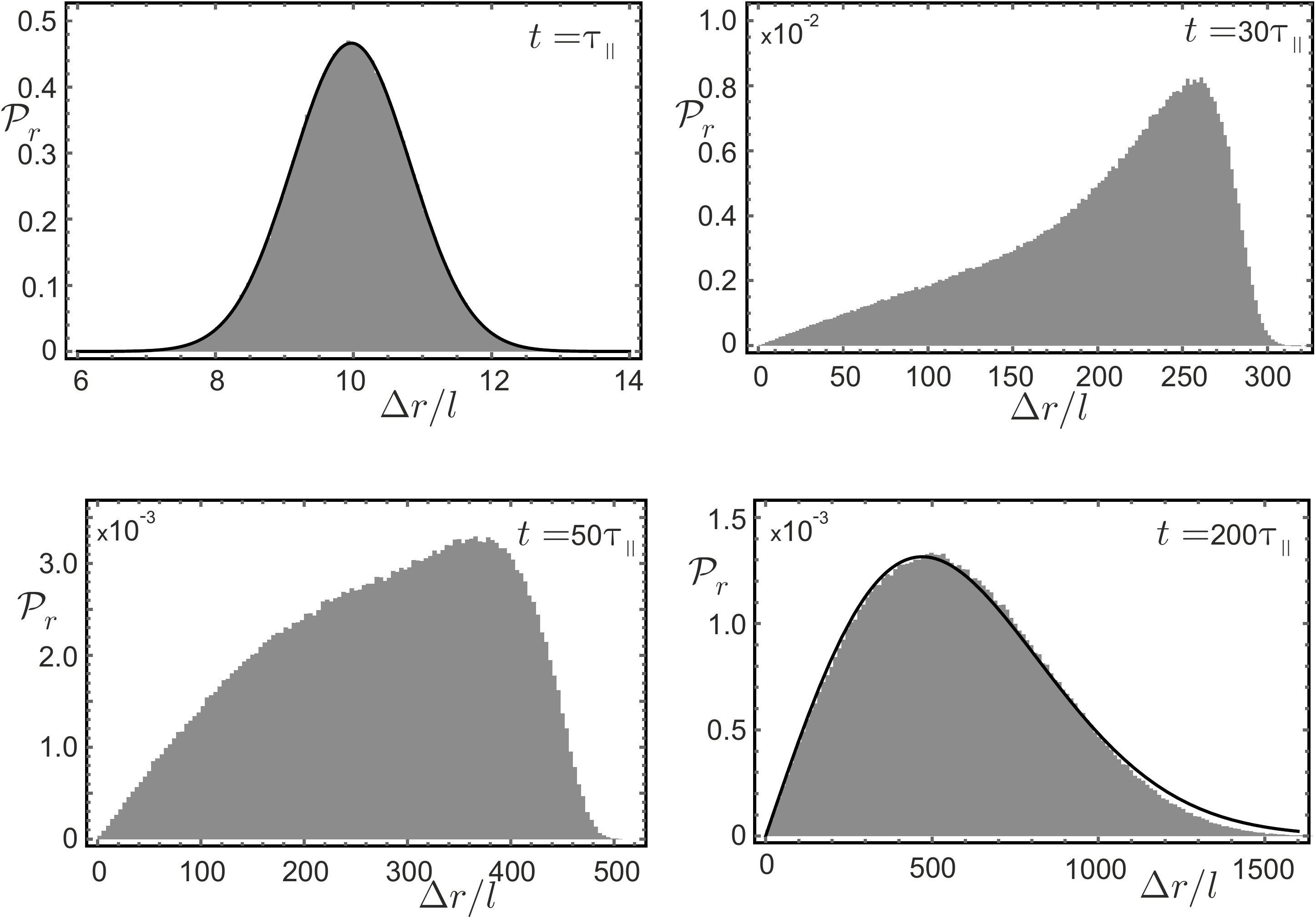}
		\caption{The probability distribution $\mathcal{P}_r$ of displacements $\Delta r$, measured
			in units of $l$, for four characteristic time values $t/\tau_{\parallel}$. Here we chose
			the same values of parameters as those used to obtain the cyan line in figure~\ref{fig6}, $P=d=10$ and $Q=w=b=1$. For $t/\tau_{\parallel}=1$ the
			solid black line represents the best fit of $\mathcal{P}_r$ to a Gaussian form. For $t/\tau_{\parallel}=200$ the solid black line is the best fit of $\mathcal{P}_r$ to the form $\mathcal{P}_r = (\Delta r/\sigma_r^2) \rme^{-\Delta r^2/(2\sigma_r^2)}$, where $\sigma_r$ is a fit parameter.}
		\label{fig10}
	\end{center}
\end{figure}

For the same parameter choice, the corresponding time evolution of the probability distribution $\mathcal{P}_r$ of radial displacement $\Delta r = \sqrt{\Delta x^2 + \Delta y^2}$ is shown in figure~\ref{fig10}. In this representation the initial two peak structure of $\mathcal{P}_x$, presented in figure~\ref{fig9}, maps onto a Gaussian centered at $\Delta r/l \approx P$. Later in time the peak of $\mathcal{P}_r$ propagates to higher values of $\Delta r/l$, and develops a shoulder for smaller displacements $\Delta r/l$ (see figure~\ref{fig10} for $t/\tau_{\parallel} = 30$). For even longer times (in our case $t/\tau_{\parallel}=50$) the shoulder becomes more pronounced and eventually the probability distribution attains the expected form $\mathcal{P}_r = (\Delta r/\sigma_r) \rme^{-\Delta r^2/(2\sigma_r^2)}$, which is typical for the diffusive regime.

\section{Conclusion}\label{conclusion}

We have studied the dynamics of a polar tracer with a concave surface in a bath consisting of active particles. By investigating the non-equilibrium statistics of the forces and torques with which the active particles push against the tracer, we were able to fully determine a set of three
effective Langevin equations for the tracer position and orientation. Thus, this procedure enabled us to reduce the complexity of the problem, by going from an involved many-body dynamics approach to a coarse-grained description of the bath, which appears in the tracer dynamics as a force drift and an exponentially correlated noise. Our effective Langevin equations contain five independent dimensionless parameters, which depend on the geometry of the tracer and the properties of active particles constituting the bath. Further work is needed in order to establish a closer connection between the parameters in our coarse-grained model and the parameters of the full many-body system
in the Brownian dynamics simulations.

Polar tracers can harness energy from the noisy non-equilibrium environment of an active bath and
thereby generate directed motion. Our work provides a complete effective description for the coupled translational and rotational tracer motion. It will help to further explore the capabilities of active baths for fueling directed transport, for example, with micro shuttles. An extension of this idea is to endow the polar tracers with some intrinsic information processing system so that they can sense their environment and act accordingly. Such smart micro shuttles can then use reinforcement learning to learn to perform some prescribed task. For example, in~\cite{schneider19} it was demonstrated how smart active particles learn to optimize their travel time in a potential landscape.

\ack

We thank Benjamin Lindner for initiating discussions. MK gratefully acknowledges financial support from the Alexander von Humboldt Foundation through a postdoctoral research fellowship.

\appendix

\section{Brownian dynamics simulations}\label{appA}

We consider a system of $N$ interacting active Brownian particles in two dimensions, which self-propel with a constant speed $v$ and have a mobility $\mu$. Their dynamics is described by
overdamped stochastic equations
\begin{eqnarray}
\dot \mathbf{r}_i = v \mathbf{u}_i - \mu \sum_{j \neq i} \nabla_{\mathbf{r}_i} V(\mathbf{r}_i - \mathbf{r}_j), \label{BDeq1} \label{rieq} \\
\dot \theta_i = \sqrt{2D_{\mathrm{R}}} \eta_i. \label{BDeq2}
\end{eqnarray}
Here $\mathbf{r}_i$ is position vector and $\mathbf{u}_i \equiv (\cos\theta_i, \sin\theta_i)$ the unit orientation vector of particle $i$, $D_{\mathrm{R}}$ denotes its rotational diffusion constant, and $\eta_i$ is Gaussian white noise of zero mean and unit variance: $\langle \eta_i \rangle = 0$, $\langle \eta_i (t) \eta_j (t') \rangle = \delta_{ij} \delta(t-t')$. We perform simulations in the regime of large $v$, which is physically most interesting. This allows us to neglect the effect of translational thermal diffusivity in (\ref{rieq}). Active particles interact with each other through pairwise forces, which are given by the negative gradient of the Weeks-Chandler-Andersen (WCA) potential
$$
V(\mathbf{r}) = \left\{
\begin{array}{ll}
4 \varepsilon \left [ \left (\frac{\sigma}{|\mathbf{r}|} \right )^{12} - \left (\frac{\sigma}{|\mathbf{r}|} \right )^{6}\right ] + \varepsilon, & \quad |\mathbf{r}| \leq 2^{1/6} \sigma, \\
0, & \quad |\mathbf{r}| > 2^{1/6} \sigma.
\end{array}
\right.
$$
Here $\varepsilon$ is the strength of the potential and $\sigma$ is the characteristic length where
the potential takes the value $\varepsilon$. We carry out simulations in a rectangular box of size $L \times L$ and use periodic boundary conditions. 

We use $\sigma$ as the unit of length, persistence time $\tau_{\mathrm{R}} = D_{\mathrm{R}}^{-1} = \sigma^2/(3\mu k_{\mathrm{B}}\mathrm{T})$ of an active particle as the unit of time, and we measure energies in units of $k_{\mathrm{B}}\mathrm{T}$, where $\mathrm{T}$ is the temperature of the solvent surrounding active particles (not to be confused with the torque $T$ used in the main text). We introduce the persistence number $P_{\mathrm{er}}=v\tau_{\mathrm{R}}/\sigma$, which measures the distance an active particle travels in approximately the same direction. The equations (\ref{BDeq1}) and (\ref{BDeq2}) can be transformed into a dimensionless form with two independent dimensionless parameters: the persistence number $P_{\mathrm{er}}$ and the potential strength $\varepsilon/k_{\mathrm{B}}\mathrm{T}$. The persistence number $P_{\mathrm{er}}$, together with the area packing fraction of active particles, $\phi=N\sigma^2\pi/(4L^2)$, determine the properties of the active bath. 

Here we consider a polar tracer immersed in the bath of interacting active particles (see figure~\ref{fig1}). We imagine our tracer as a semicircle of radius $R$ composed of particles having effective diameter $\sigma$. Then, an active particle interacts with a particle of the 
semicircle through a repulsive contact force, derived from the WCA potential, provided that the distance between them is smaller than $2^{1/6}\sigma$. The position of the polar tracer is described by the coordinates of its center of mass, $\mathbf{r} = (x,y)$, and the angle its symmetry axis makes with the $x$-axis of the lab frame (figure 1a). Now the equations of motion of the tracer can be written in the form
\begin{eqnarray}
\dot x = \mu_{\parallel}F_{\parallel}\cos\theta - \mu_{\perp}F_{\perp}\sin\theta, \\
\dot y = \mu_{\parallel}F_{\parallel}\sin\theta + \mu_{\perp}F_{\perp}\cos\theta, \\
\dot \theta = \kappa T.\label{BDeq5}
\end{eqnarray}
Here, $F_{\parallel}$ and $F_{\perp}$ are the projections on $\mathbf{e}_{\parallel}$ and $\mathbf{e}_{\perp}$ of the resulting force exerted by active particles on the tracer, and similarly $T$ is the projection on the unit vector $\mathbf{e}_z = \mathbf{e}_{\parallel} \times \mathbf{e}_{\perp}$ of the resulting torque on the tracer. The translational mobilities of the tracer are denoted by $\mu_{\parallel}$ and $\mu_{\perp}$, while its rotational mobility is denoted by $\kappa$. 

The number of active particles is fixed to $N=10^4$, and the area $L^2$ of the simulation box is adjusted to obtain the required packing fraction $\phi$. We set $\varepsilon/k_{\mathrm{B}}\mathrm{T}=100$, $R/\sigma = 5$, $\mu_{\parallel}/\mu = 0.2$, $\mu_{\perp}/\mu = 0.1$ and $\kappa\sigma^2/(3\mu) = 10^{-3}$. Equations (\ref{BDeq1}) - (\ref{BDeq5}) are integrated using a simple Euler scheme with a time step of $\delta t/\tau_{\mathrm{R}} = 10^{-5}$. The simulation time goes up to $t/\tau_{\mathrm{R}}=5000$, and all results are averaged over 15 independent simulation runs. 

\begin{figure}[htb]
	\begin{center}
		\includegraphics[width=13cm]{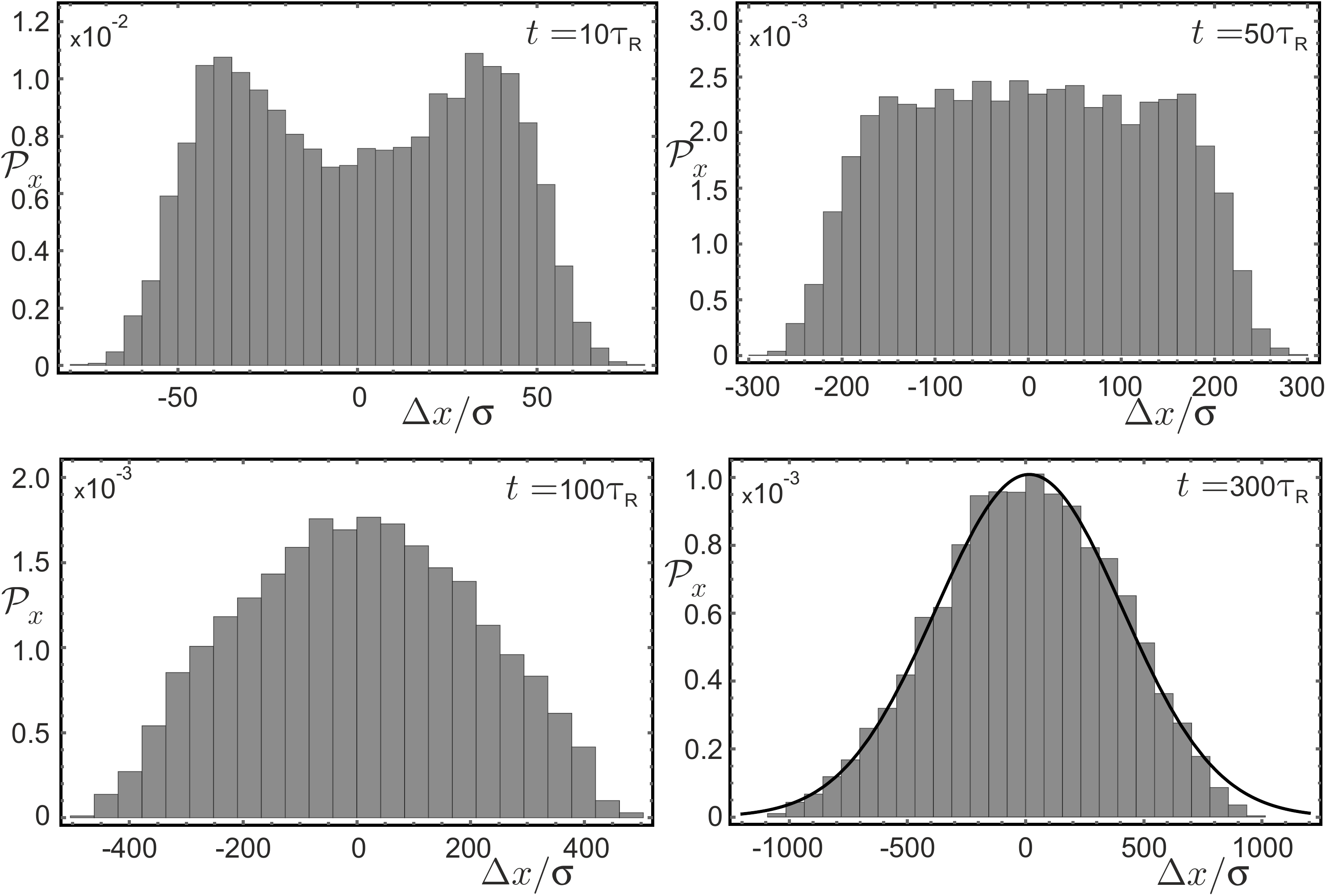}
		\caption{The probability distribution $\mathcal{P}_x$ of tracer displacements $\Delta x$, measured in units of $\sigma$, for four characteristic time values $t/\tau_{\mathrm{R}}$. The persistence number takes value $P_{\mathrm{er}} = 80/3$ and the packing fraction is chosen to be $\phi \approx 0.08$. For $t/\tau_{\mathrm{R}}=300$ the solid black line provides the best fit of $\mathcal{P}_x$ to a Gaussian form.}
		\label{figA1}
	\end{center}
\end{figure}

A typical snapshot from our Brownian dynamics simulation is presented in figure~\ref{fig1}b. The probability distributions of $F_{\parallel}$, $F_{\perp}$ and $T$ and their time auto-correlation functions are shown in figure~\ref{fig2}, while the cross-correlation function $\langle F_{\perp}(t_0) T(t_0+t) \rangle$ obtained in this approach is presented in figure~\ref{fig3}. Finally, in figure~\ref{figA1} we give the probability distributions $\mathcal{P}_x$ of tracer displacement $\Delta x$ for several characteristic times.
\setcounter{section}{1}

\end{document}